\documentclass[10pt,twocolumn,letterpaper]{IEEEtran}
\usepackage{booktabs}
\usepackage{graphics}
\usepackage{amsmath}
\usepackage{xcolor}
\usepackage{dsfont}
\usepackage{graphicx}
\usepackage[export]{adjustbox}

\DeclareMathOperator*{\minimize}{minimize}
\renewcommand \baselinestretch{0.98}
\usepackage{fullpage}
\usepackage{amssymb,amsmath}\newcommand{\RR}{\mathbb R}
\usepackage{morefloats}

\usepackage{subcaption}

\IEEEoverridecommandlockouts                              

\title{\LARGE \bf
A Graph-constrained Changepoint Detection Approach for ECG Segmentation
}

\author{Atiyeh Fotoohinasab,  Toby Hocking, and Fatemeh Afghah,
\thanks{A. Fotoohinasab, T. Hocking and F. Afghah are with the School of Informatics, Computing and Cyber Systems at Northern Arizona University.}
}

  \renewcommand{\baselinestretch}{.92}
\begin{document}

\maketitle
\thispagestyle{empty}
\pagestyle{empty}

\begin{abstract}

Electrocardiogram (ECG) signal is the most commonly used non-invasive tool in the assessment of cardiovascular diseases. Segmentation of the ECG signal to locate its constitutive waves, in particular the R-peaks, is a key step in ECG processing and analysis. Over the years, several segmentation and QRS complex detection algorithms 
have been proposed with different features; however, their performance highly depends on applying preprocessing steps which makes them unreliable in real-time data analyzing of ambulatory care settings and remote monitoring systems, where the collected data is highly noisy. Moreover, some issues are still remained with the current algorithms in regard to the diverse morphological categories for the ECG signal and their high computation cost. In this paper, we introduce a novel graph-based optimal changepoint detection (GCCD) method for reliable detection of R-peak positions without employing any preprocessing step. The proposed model guarantees to compute the globally optimal changepoint detection solution. It is also generic in nature and can be applied to other time-series biomedical signals. Based on the MIT-BIH arrhythmia (MIT-BIH-AR) database, the proposed method achieves overall sensitivity Sen = 99.76, positive predictivty PPR = 99.68, and detection error rate DER = 0.55 which are comparable to other state-of-the-art approaches\footnote{This material is based upon work supported by the National Science Foundation under Grant Number 1657260.}.

\end{abstract}


\section{INTRODUCTION}
The electrocardiogram (ECG) is a quasi-periodical, rhythmically repeating biomedical signal that includes information about the electrical activities of cardiac muscle. One cycle of ECG, called a heartbeat, is delineated by arrangements of P, QRS-complex and T waveforms as well as PQ and ST segments. Out of all the waveforms, the QRS-complex is the most striking waveform as it represents the ventricular depolarization and reflects the major part of the electrical activity of the heart. Precisely detecting the location of R-peak as the highest and the only positive peak within the QRS complex plays a critical role in obtaining the morphology of the QRS-complex. Besides, the localization of R-peak serves as the basis for the automated determination of the heart rate and thereby, it is a significant criterion for discerning any heart abnormality. Not only heart diseases but also many other diseases can be diagnosed in a non-invasive way using R-peak detection due to the relationship between heart rate variability (HRV) and several physiological systems (vasomotor, respiratory, central nervous, thermoregulatory, etc.). Revealing R-peaks is a simple task in noise-free ECG signals; however, the challenge emerges when the ECG signal is corrupted by noise and artifacts.

Several methods have been proposed to detect different waves in the ECG signal.
However, the majority of these methods still suffer from several drawbacks such as misdetection of ECG waveforms with determinant morphological patterns in some life-threatening heart arrhythmias, or the high complexity of the algorithm that is a barrier for online processing or long-recording analysis. Typically, current methods consist of two main steps; pre-processing, and detection. In the pre-processing step, the algorithm attempts to eliminate the noise and artifacts, and highlight the relevant sections of the ECG \cite{castells2015simple, sharma2017qrs}. In the second step, most of the methods locate R-peaks using the result from the pre-processing step in the first place, then other waves are detected by defining a set of heuristic rules \cite{hou2018real}. 
However, in real-time data processing of ambulatory care settings and remote monitoring systems, where the collected data is highly noisy, preprocessing-based algorithms are less effective. Besides, the performance of most detection algorithms depends highly on the used dataset, and also, incorrect detection of R-peak leads to the incorrect identification of other waves. 

Most of the state-of-the-art methods in the field of ECG waves detection are either based on the wavelet transform, Hidden Markov models (HMM), and simple mathematical operations such as differentiation, integration, and squaring. Wavelet transform is a common approach to capture non-stationary behaviors of the ECG signal \cite{park2017r}. 
However, the caveat with the Wavelet transform is the difficulty in determining a proper mother wavelet derived from the various shape of the QRS-complex and finding the required threshold in the detection step. More importantly, discrete wavelet transforms (DWT) often fail to derive reliable results from short-recordings of the signal. Algorithms stem from simple mathematical operations are computationally efficient, and thereby ideal for real-time analysis of the large data sets. Pan and Tompkins \cite{PanTompkins1985} algorithm and its modifications \cite{gutierrez2015novel} 
are the most popular techniques in this category. However, their high performance is guaranteed in case of the ECG signal is weakly disturbed by the noise. Hidden Markov models (HMMs) are also graphical models that have highly drawn attention in the task of ECG segmentation in order to consider the temporal dependency among the waveforms. In addition to the aforementioned methods, some studies addressed the problem of ECG delineation using deep learning. Although deep learning-based algorithms demonstrated high performance in the classification problems, they rely on large scale datasets for training the algorithm and suffer from the imbalanced class problem \cite{mousavi2020single, mousavi2020han, mousavi2019inter, mousavi2019ecgnet, mousavi2019sleepeegnet, parviziomran2019data}.

In this work, we leverage a graph-constrained changepoint detection (GCCD) model introduced by Hocking et al. in \cite{hocking2017log} in order to detect R-peak positions in a fast and effective way. Hocking et al. \cite{hocking2017log} proposed an algorithm that can solve changepoint detection problems with graph constraints on segment's means with log-linear complexity in the amount of data, thereby has high efficiency on huge data sets. Their model can represent a time series signal as a set of constant value segments of varying lengths, as well as their precise locations over space or time. They have shown that their model has high accuracy in detecting abrupt changes in genomic data such as DNA copy number profiles \cite{HOCKING-breakpoints} and ChIP-sequencing \cite{HOCKING-PeakSeg}. In this paper, we indicate that this model not only is applicable for a periodic signal but also can outperform other state-of-the-art methods concerned with ECG delineation. 

There are a few studies in the literature that have exploited changepoint detection models to find abrupt changes in heart rhythm. However, up to our best knowledge, this is the first study where a changepoint detection model is used to segment ECG waveforms. One of the key contributions of this work is using graphs to some extent employ biological prior knowledge of the signal into the model and lessen the complexity of the model. The proposed method does not need a pre-processing step since it leverages the sparsity of changepoints to denoising the signal as well as detecting abrupt changes. As a result, it can be more effective in real-time processing of noisy data. Besides, GCCD takes the whole cardiac beat into account to detect and study each wave separately, as opposed to most of the current methods which use the location of R-wave to detect other waves. 

The rest of the paper is organized as follows. In the next section, we describe the proposed Graph-Constrained Changepoint Detection model and its application in localization of R-peak. Section \ref{sec:Experimental Studies} provides a description of the dataset used in this study and a discussion of the results as well as a comparison between the performance of the proposed algorithm and other state-of-the-art algorithms. Finally, \ref{sec:CONCLUSIONS} summarizes this research work and its contribution.

\section{Methodology}
 \label{sec:GCCD}
 
\begin{figure*}[htb]
      \centering
      \includegraphics[width=\linewidth,height=\textheight,keepaspectratio]{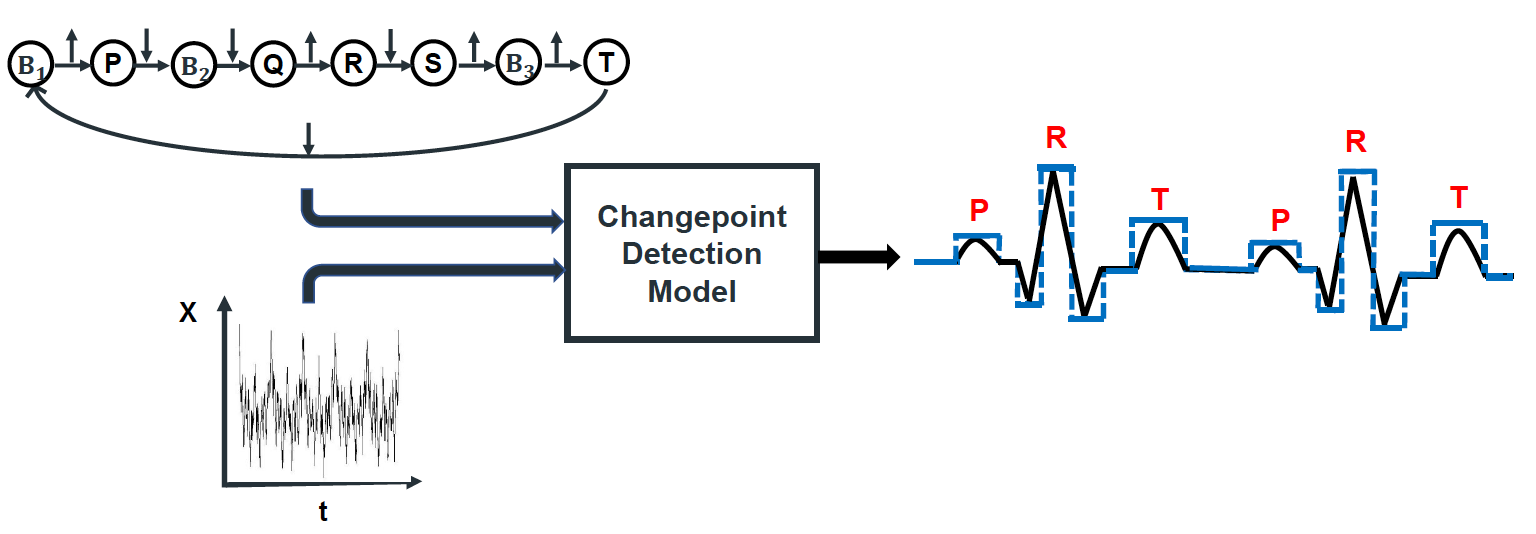}
      \caption{An overview of the Graph-Constrained Changepoint Detection (GCCD) model. The GCCD model takes a constraint graph and a raw ECG signal as inputs, and then, detects segments corresponding to the nodes of the constraint graph at the output. The up/down arrows above each edge represents the expected up/down change in the segment means, respectively. In this model, the proposed constraint graph has a vertex for each main waveform (i.e. $P, Q, R, S, T$) as well as Baselines ($B_1$-$B_3$), which are intermediate states.}
      \vspace{-15pt}
      \label{System_model}
   \end{figure*}

The proposed method treats the problem of ECG waveforms delineation as a changepoint detection problem over a non-stationary ECG signal. It represents the periodic ECG signal as piece-wise locally stationary pieces so as each piece is the mean of one segment of data points. The model takes a raw ECG signal and a predefined graph, which encodes expected changepoints from the signal, as inputs and yields the onset/offset and the mean of desired segments. Figure \ref{System_model} illustrates an overview of the proposed algorithm in detecting waveforms of a normal ECG. In the next sections, we will briefly describe the GCCD model.

\subsection{Graph-Constrained Changepoint Detection Model}

 Let us assume a constrained changepoint model represented with a graph $G=(V,E)$, where the vertices $V\in\{1,\dots,|V|\}$ represent the hidden states or segments (not necessarily a waveform), and the directed edges $E\in\{1,\dots,|E|\}$ represent the possible changes between the states or segments. Each edge $e\in E$ encodes the following associated data based on the prior knowledge of expected sequences of changes:
 \begin{itemize}
    \item The source $\underline v_e\in V$ and target $\overline v_e\in V$ vertices/states for a changepoint $e$ from $\underline v_e$ to $\overline v_e$.
    \item A non-negative penalty constant $\lambda_e\in\RR_+$ which is the cost of changepoint $e$.
    \item A constraint function $g_e:\RR\times\RR\rightarrow\RR$ which defines the possible mean values before and after each changepoint $e$. If $m_i$ is the mean before the changepoint and $m_{i+1}$ is the mean after the changepoint, then the constraint is $g_e(m_i,m_{i+1})\leq 0$. These functions can be used to constrain the direction (up or down) and/or the magnitude of the change (greater/less than a certain amount).
\end{itemize}

Mathematically, given the input signal $Y=\{y_1,\dots,y_n\}$ and the graph $G=(V,E)$, finding changepoints ~$\mathbf c$, segment means $\mathbf m$, and hidden states $\mathbf s$ can be executed by solving the following optimization problem:

\begin{align}
  \label{eq:op-c}
  & \minimize_{
  \substack{
  \mathbf m\in\RR^N,\, \mathbf s\in V^N\\
   \mathbf c\in\{0,1,\dots,|E|\}^{N-1}\\
  }
  } \ \ 
      \sum_{i=1}^N \ell(m_i, z_i) + \sum_{i=1}^{N-1} \lambda_{c_i} \\
        \text{s. t\ \ } &\ \text{no change: }c_i = 0 \Rightarrow m_i=m_{i+1} ~\& ~s_i=s_{i+1} \, \label{eq:nochange-constraint}\\ \nonumber
    &\ \text{change: }c_i \neq 0 \Rightarrow g_{c_i}(m_i,m_{i+1})\leq 0 ~~\& \\
        &(s_i,s_{i+1})=(\underline v_{c_i},\overline v_{c_i}).\label{eq:change-constraint}
\end{align}

The above equation consists of a data-fitting term involving the negative log-likelihood $\ell$ of each data point, and a model complexity term involving the cost of each changepoint $\lambda_{c_i}$. More specifically, the positive penalty parameter $\lambda$ regularizes the number of predicted changepoints/segments by the model. 
A larger penalty $\lambda$ yields a changepoint vector $\mathbf c$ which is more sparse, or more specifically, reduces the number of changepoints. 
The constraint function $g_e$ also encodes the expected up/down change and the least amplitude gap between the mean of two states. For all possible changes $i$ such that $c_i=0$, the constraint~(\ref{eq:nochange-constraint}) enforces no change in mean $m_i=m_{i+1}$ and state $s_i=s_{i+1}$ variables. For all possible changes $i$ such that $c_i\neq 0$, the constraint~(\ref{eq:change-constraint}) enforces a change in mean specified by the constraint function $g_{c_i}(m_i,m_{i+1})\leq 0$, and a change in state $(s_i,s_{i+1})=(\underline v_{c_i},\overline v_{c_i})$. 
Besides, the above optimization problem exploits the dynamic programming algorithm to recursively compute cost functions over the data points and hidden states.

\subsection{ECG Waveforms Detection}
\label{sec:ECG segmentation}

At first, the topology of the constraint graph $G$ and its corresponding parameters are defined based on the different possible morphologies for each waveform (P-, QRS-, T-waveforms, etc.) over one heart cycle. This constraint graph specifies biological prior knowledge about the expected morphological changes within one cycle of ECG. More specifically, each node in the constraint graph serves an expected hidden state/segment in the signal, and each edge encodes the required conditions for the transition to a new state. These conditions are determined based on the expected minimum amplitude difference of two successive segments and the polarity of a waveform. Then, the GCCD model segments the raw ECG signal using the prior knowledge provided by the constraint graph. It is worth re-emphasizing that any preprocessing step is no longer required with the GCCD model.

\section{Experimental Studies}
\label{sec:Experimental Studies}
\subsection{Dataset}

We evaluated the GCCD model using the MIT-BIH arrhythmia (MIT-BIH-AR) database consists of 48 ECG recordings taken from 47 subjects, and each recording is sampled at 360 Hz and 30 min in length with 200 mV amplitude resolution \cite{moody2001impact, goldberger2000physiobank}. Each recording is composed of two ambulatory ECG channels from the modified lead II (MLII) and one of the modified leads V1, V2, V4, or V5. In this study, only the MLII was used, and thereby the records 102 and 107 were excluded from consideration. The database contains annotations of both heartbeat class information and R-wave positions information verified by two or more expert cardiologists independently. We employed the provided annotations for R-wave positions to evaluate the performance of the proposed algorithm.

\subsection{Experimental Results}
\label{sec:Results}
\begin{figure*}[htb]
  \begin{subfigure}[t]{0.5\textwidth}
    \centering\includegraphics[width=\textwidth]{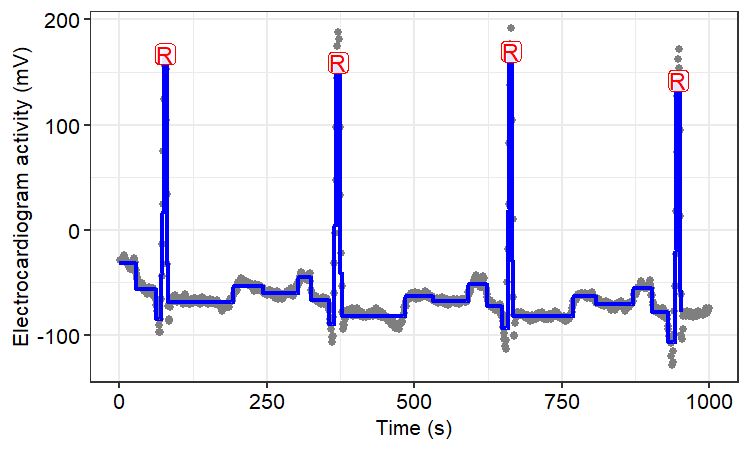}
    \caption{}
  \end{subfigure}
    \begin{subfigure}[t]{0.5\textwidth}
    \centering\includegraphics[width=\textwidth]{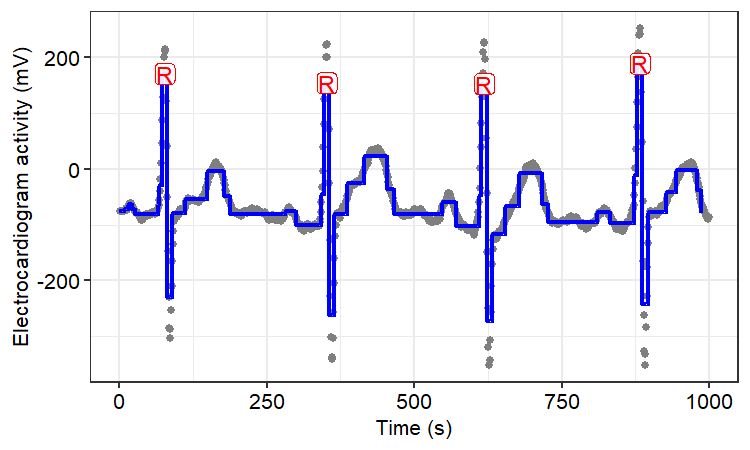}
    \caption{}
  \end{subfigure}
  \begin{subfigure}[t]{0.5\textwidth}
    \centering\includegraphics[width=\textwidth]{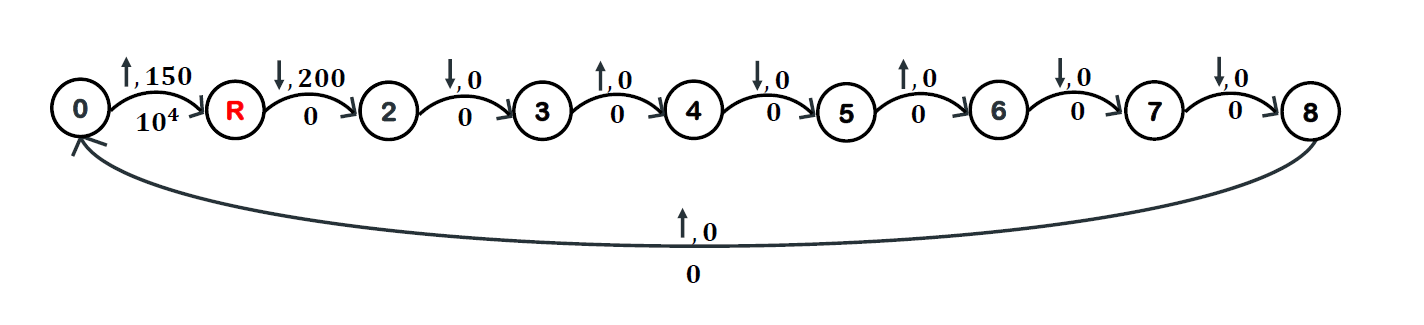}
    \caption{}
  \end{subfigure}
  \begin{subfigure}[t]{0.5\textwidth}
    \centering\includegraphics[width=\textwidth]{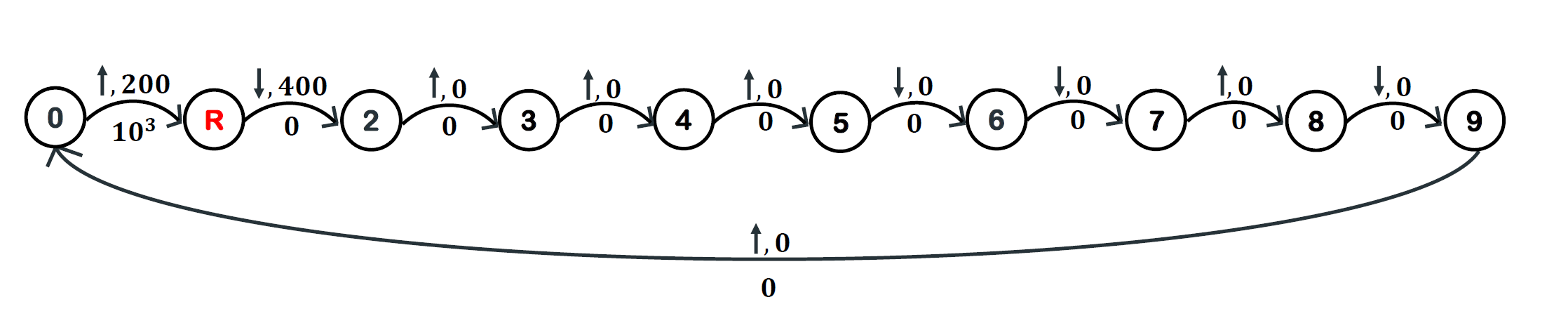}
    \caption{}
  \end{subfigure}
  \caption{The demonstration of R-peak detection using the proposed model for records 100 and 230 from the MIT-BIH-AR dataset.
    \textbf{a-b:} The proposed model represents records of 100 and 230, respectively, as piece-wise locally stationary segments (the blue lines). Extracted R-peak positions are marked as red labels.
    \textbf{c-d:} The graph structure for above two proposed models. Each graph has a vertex for each state including state ``R" for the R-wave. Below each edge $e$ we show the penalty $\lambda_e$, which is either a constant $\lambda>0$ or zero; above we show the constants $\delta,\gamma$ in the constraint function $g_e(m_i,m_{i+1})= \delta(m_i - m_{i+1})+\gamma\leq 0$ ($\delta=1$ for a non-decreasing change which is shown with $\uparrow$, $\delta=-1$ for a non-increasing change which is shown with $\downarrow$, $\gamma \geq 0$ is the minimum magnitude of change).}\label{fig:ECG_result}
\end{figure*}

Fig. \ref{fig:ECG_result} demonstrates an example of  the performance of this proposed model in the localization of R-peak for a window of two records (\# 100 and \#230) from the MIT-BIH-AR dataset. The performance of R-peak detection algorithms is usually evaluated in terms of sensitivity (Sen), positive predictivity rate (PPR) and detection error rate (DER), which are calculated by:

\begin{align}
\label{eq:sen}
Sen (\%) =  \frac{TP}{TP+FN}\times{100}\\
\label{eq:PPR}
PPR (\%) =  \frac{TP}{TP+FP}\times{100}\\
\label{eq:DER}
DER (\%) =  \frac{FN+FP}{TP+FN}\times{100}
\end{align}

In order to compare the performance of the proposed method against other methods in the literature, we adopted the above statistical metrics for the evaluation of the GCCD algorithm. Table \ref{tab:Performance_evaluation} represents the R-peak detection success for GCCD algorithm considering all 48 records in the MIT-BIH dataset. As the table shows, the proposed method achieves notable results, Sen = 99.76, PPR = 99.68, and DER = 0.55, in R-peak detection. 
The constraint graph for each record is defined manually by matching the overall morphological properties of the signal to pre-specified categories for each waveform \cite{tafreshi2014automated}. 
All the Sen and PPR values are higher than 99\% except 6 cases. For 26 cases of detection in the dataset, Sen and PPR values are 100\% separately. Records where the Sen and PPR values are less than 99\% are mainly due to two reasons: in some cycles of the record, the morphology of the QRS-complex possesses no R-peak, or the amplitude of the R-peak is too low compared to the amplitude of its next negative peak; S wave. It worth to mention that the results for records 105 and 108, as the noisiest records in the data set, demonstrate the capability of the model in the detection of R-peak locations in the presence of high noise level without using a preprocessing step for removing the noise. Table \ref{tab:Comparison_of_methods} compares the performance of the GCCD algorithm against other state-of-the-art methods. Results in this table present that the performance of the proposed model without using any preprocessing step is comparable to the other R-wave detection methods in the literature using a preprocessing step.

\begin{table}[ht]
\caption{Performance evaluation of the proposed method using the MIT-BIH-AR database}
 \centering{
\label{tab:Performance_evaluation}
\resizebox{1.\linewidth}{!}{ 
\begin{tabular}{cccccccc}
\toprule
\textbf{Record} & \textbf{Total beats} & \textbf{TP} & \textbf{FP} & \textbf{FN} & \textbf{Sen (\%)} & \textbf{PPR (\%)} & \textbf{DER (\%)}\\ 
\midrule
\texttt 100 & 2273 & 2273 & 0 & 0 & 100 & 100 & 0 \\
\texttt 101 & 1865 & 1865 & 0 & 0 & 100 & 100 & 0 \\
\texttt 102 & 2187 & 2187 & 0 & 0 & 100 & 100 & 0 \\
\texttt 103 & 2084 & 2084 & 0 & 0 & 100 & 100 & 0 \\
\texttt 104 & 2229 & 2572 & 24 & 5 & 99.77 & 98.93 & 1.3 \\
\texttt 105 & 2572 & 2572 & 0 & 0 & 100 & 100 & 0 \\
\texttt 106 & 2027 & 2022 & 0 & 5 & 99.75 & 100 & 0.24 \\
\texttt 107 & 2137 & 2137 & 0 & 0 & 100 & 100 & 0\\
\texttt 108 & 1765 & 1765 & 2 & 0 & 100 & 99.88 & 0.11 \\
\texttt 109 & 2532 & 2532 & 0 & 0 & 100 & 100 & 0 \\
\texttt 111 & 2124 & 2120 & 5 & 4 & 99.76 & 99.81 & 0.42 \\
\texttt 112 & 2539 & 2539 & 0 & 0 & 100 & 100 & 0 \\
\texttt 113 & 1795 & 1795 & 0 & 0 & 100 & 100 & 0 \\
\texttt 114 & 1879 & 1864 & 34 & 15 & 99.20 & 98.20 & 2.6 \\
\texttt 115 & 1953 & 1535 & 0 & 0 & 100 & 100 & 0 \\
\texttt 116 & 2412 & 2410 & 1 & 2 & 99.95 & 99.91 & 0.12 \\
\texttt 117 & 1535 & 1535 & 0 & 0 & 100 & 100 & 0 \\
\texttt 118 & 2278 & 2276 & 4 & 2 & 99.91 & 99.82 & 0.26 \\
\texttt 119 & 1987 & 1987 & 0 & 0 & 100 & 100 & 0 \\
\texttt 121 & 1863 & 1863 & 0 & 0 & 100 & 100 & 0 \\
\texttt 122 & 2476 & 2476 & 0 & 0 & 100 & 100 & 0 \\
\texttt 123 & 1518 & 1518 & 0 & 0 & 100 & 100 & 0 \\
\texttt 124 & 1619 & 1617 & 10 & 2 & 99.87 & 99.38 & 0.74 \\
\texttt 200 & 2601 & 2521 & 50 & 80 & 96.92 & 98.05 & 4.99 \\
\texttt 201 & 1963 & 1962 & 9 & 1 & 99.94 & 99.54 & 0.50 \\
\texttt 202 & 2136 & 2134 & 7 & 2 & 99.90 & 99.67 & 0.42 \\
\texttt 203 & 2980 & 2934 & 43 & 46 & 98.45 & 98.55 & 2.98 \\
\texttt 205 & 2656 & 2656 & 0 & 0 & 100 & 100 & 0 \\
\texttt 207 & 1860 & 1812 & 81 & 48 & 97.41 & 95.72 & 6.99 \\
\texttt 208 & 2955 & 2946 & 2 & 9 & 99.69 & 99.93 & 0.37 \\
\texttt 209 & 3005 & 3005 & 0 & 0 & 100 & 100 & 0 \\
\texttt 210 & 2650 & 2648 & 32 & 2 & 99.92 & 98.80 & 1.28 \\
\texttt 212 & 2748 & 2748 & 0 & 0 & 100 & 100 & 0 \\
\texttt 213 & 3251 & 3249 & 3 & 2 & 99.93 & 99.90 & 0.15 \\
\texttt 214 & 2262 & 2262 & 10 & 0 & 100 & 99.55 & 0.44 \\
\texttt 215 & 3363 & 3361 & 6 & 2 & 99.94 & 99.82 & 0.23 \\
\texttt 217 & 2208 & 2208 & 0 & 0 & 100 & 100 & 0 \\
\texttt 219 & 2154 & 2154 & 0 & 0 & 100 & 100 & 0 \\
\texttt 220 & 2048 & 2048 & 0 & 0 & 100 & 100 & 0 \\
\texttt 221 & 2427 & 2425 & 3 & 2 & 99.91 & 99.87 & 0.20 \\
\texttt 222 & 2483 & 2472 & 0 & 11 & 99.55 & 100 & 0.44 \\
\texttt 223 & 2605 & 2604 & 3 & 1 & 99.96 & 99.88 & 0.15 \\
\texttt 228 & 2053 & 2050 & 5 & 3 & 99.85 & 99.75 & 0.38 \\
\texttt 230 & 2256 & 2256 & 0 & 0 & 100 & 100 & 0 \\
\texttt 231 & 1571 & 1571 & 0 & 0 & 100 & 100 & 0 \\
\texttt 232 & 1780 & 1779 & 1 & 1 & 99.94 & 99.94 & 0.11 \\
\texttt 233 & 3079 & 3063 & 11 & 16 & 99.48 & 99.64 & 0.87 \\
\texttt 234 & 2753 & 2753 & 0 & 0 & 100 & 100 & 0 \\
\bottomrule 
\texttt Total & 109,494 & 109,148 & 346 & 261 & 99.76 & 99.68 & 0.55\\
\end{tabular}}}
\end{table}

\begin{table}[ht]
\caption{Comparison of performance of several R-peak detection methods using the MIT-BIH-AR database}
 \centering{
\label{tab:Comparison_of_methods}
\resizebox{1.\linewidth}{!}{ 
\begin{tabular}{ccccccc}
\toprule
\textbf{Method} & \textbf{TP} & \textbf{FP} & \textbf{FN} & \textbf{Sen (\%)} & \textbf{PPR (\%)} & \textbf{DER (\%)}\\ 
\midrule
\texttt Pan and Tompkins \cite{PanTompkins1985} &  115,860 & 507 & 277 & 99.76 &  99.56 & 0.68\\
\texttt Park et al. \cite{park2017r} & 109,415 & 99 & 79 & 99.93 & 99.91 & 0.163\\
\texttt Chen et al. \cite{chen2017qrs} & 109,250 & 203 & 193 & 99.82 &  99.81 &  0.36\\
\texttt Farashi \cite{farashi2016multiresolution} & 109,692 &  163 & 273 & 99.75 & 99.85 & 0.40 \\
\texttt Sharma and Sunkaria \cite{sharma2017qrs} & 109,381 & 136 & 113 & 99.50 & 99.56 & 0.93\\
\texttt Castells-Rufas and Carrabina \cite{castells2015simple} & 108,880 &  353 & 614 & 99.43 & 99.67 & 0.88\\
\texttt Yazdani and Vesin \cite{yazdani2016extraction} & 109,357 &  108 & 137 &  99.87 &  99.90 & 0.22\\
\textbf {GCCD Model} & \textbf {109,148} &  \textbf{346} & \textbf {261} &  \textbf {99.76} &  \textbf {99.68} & \textbf {0.55}\\
\hline
\end{tabular}}}
\end{table}

The results provided in this research work justifies that changepoint detection models are promising approaches for the detection of ECG waveforms. The proposed model has the ability to detect R-peak positions for various morphologies of the ECG. Additionally, GCCD model can potentially detect other waveforms than R wave (i.e, P,Q,S, and T waves) by including their corresponding knowledge to the constraint graph. In this research work, we only detect R-peak positions since the MIT-BIH-AR database has just provided with the annotations for this peak. As mentioned before, for some cases in the database, the performance of the GCCD model is notably low due to the absence of the R-peak or the low amplitude of this peak in some cycles. The efficiency of the GCCD model for these specific cases can be enhanced by using a multi-path constraint graph. A multi-path constraint graph is useful in identifying all ECG waveforms as well as QS-complex possessing no R-peak. Additionally, the constraint graph can be learned in a batch mode based on each record to surmise the model from the manual definition of the constraint graph, which is considered as future work.

\section{Conclusion}
\label{sec:CONCLUSIONS}
The detection of ECG waveforms plays a paramount role in most of the automated ECG analysis tools. The most common features found in the literature for the detection of several heart diseases are computed from the morphological characteristics of R-peak, as the most important fiducial point within one cycle of the ECG signal. In this paper, we applied for the first time a changepoint detection model, named graph-constrained changepoint detection model, in order to extract the positions of R-peak. The constraint graph encodes prior knowledge of expected changepoints per heart cycle based on the possible morphologies for the ECG signal. The proposed model can detect R-peak locations in a log-linear complexity and requires no preprocessing steps. The evaluation results demonstrated the proposed method has a high performance for the detection of R-peak locations, especially in noisy records, compared to the existing algorithms in the literature while their performance highly relies on removing noise in the preprocessing step.



\addtolength{\textheight}{-12cm}   


\bibliographystyle{plain}
\bibliography{references}
\end{document}